\documentclass[aps,pra,preprint,superscriptaddress,longbibliography]{revtex4-2}

\usepackage{amsmath}
\usepackage{amsthm}
\usepackage{amsfonts}
\usepackage{amssymb}
\usepackage{xcolor,graphicx}
\usepackage{tikz}
\usepackage{color}
\usepackage[pdftex]{hyperref} 
\usepackage{longtable}
\usepackage{array}
\usepackage{dsfont}

\begin{document}
	
	\title{Optomechanical simulation of a parametric oscillator}
	
	\author{F.~E. Onah}
	\email[e-mail: ]{a00834081@itesm.mx}
	\affiliation{Tecnol\'ogico de Monterrey, Escuela de Ingenier\'ia y Ciencias, Ave. Eugenio Garza Sada 2501, Monterrey, N.L., Mexico, 64849}
	\affiliation{The Division of Theoretical Physics, Physics and Astronomy, University of Nigeria Nsukka, Nsukka Campus, Enugu State, Nigeria}

	\author{C. Ventura-Vel\'azquez}
	\email[e-mail: ]{christian.venturavelazquez@upol.cz}
	\affiliation{Deparment of Optics, Palack\'y University, 17.Listopadu 12, 77146 Olomouc, Czech Republic}

	\author{F.~H. Maldonado-Villamizar}
	\email[e-mail: ]{fmaldonado@inaoep.mx}
	\affiliation{CONACYT - Instituto Nacional de Astrof\'isica, \'Optica y Electr\'onica, Calle Luis Enrique Erro No. 1. Sta. Ma. Tonantzintla, Pue. C.P. 72840, Mexico}

	\author{B.~R. Jaramillo-\'Avila}
	\email[e-mail: ]{jaramillo@inaoep.mx}
	\affiliation{CONACYT - Instituto Nacional de Astrof\'isica, \'Optica y Electr\'onica, Calle Luis Enrique Erro No. 1. Sta. Ma. Tonantzintla, Pue. C.P. 72840, Mexico}

	\author{B.~M. Rodr\'iguez-Lara}
	\email[e-mail: ]{bmlara@tec.mx}
	\affiliation{Tecnol\'ogico de Monterrey, Escuela de Ingenier\'ia y Ciencias, Ave. Eugenio Garza Sada 2501, Monterrey, N.L., Mexico, 64849}
	
	\date{\today}
	
	\begin{abstract}
	We study an optomechhanical device supporting at least three optical modes in the infrared telecommunication band and three mechanical vibration modes. 
	We model the coherent driving of each optical mode, independently of each other, to obtain an effective Hamiltonian showing the different types of parametric processes allowed in the device. 
	We propose a bichromatic driving scheme, in the lossy optical cavity regime, under a mean field approximation, that provides the quantum simulation of a parametric oscillator with optical control of its parameters.
	\end{abstract}

	\maketitle
	
\section{Introduction}

Cavity optomechanics explores the interaction between optical and mechanical excitation modes in the quantum regime \cite{Aspelmeyer2014}. 
In an optomechanical device, the interaction arises from the radiation pressure that optical modes exert on a material structure and, in consequence, its mechanical vibration modes. 
The structure is typically a solid and its quantum control is in itself a milestone in physics research \cite{Aspelmeyer2014}.
Further motivation for research in the field comes from quantum technologies, where photon-to-photon transducers \cite{Guha2021,Balram2022} and quantum memories \cite{Wallucks2020} are necessary for quantum communication and information processing, and quantum sensing, where optomechanical devices take the role of high precision sensors \cite{Qiao2018,Li2021}, for example.

There exists a plethora of experimental platforms to realize cavity optomechanics; for example, suspended micromirrors \cite{Groeblacher2009}, microtoroids \cite{Riviere2011}, microsphere resonators \cite{Ma2007}, micromechanical membranes in a superconducting microwave cavity \cite{Teufel2011}, one-dimensional photonic crystal nanobeams \cite{Chan2011}, and levitated nanoparticles \cite{Delic2020}. 
These devices vary in size and shape, covering a wide range of frequencies and coupling strength values.
However, they are accurately modeled by a relatively simple Hamiltonian, 
\begin{eqnarray}
	\frac{\hat{H}}{\hbar} = \omega \hat{a}^{\dagger} \hat{a} + \omega_{m} \hat{b}^{\dagger} \hat{b} - g_{0} \hat{a}^{\dagger} \hat{a} \left( \hat{b}^{\dagger} + \hat{b}  \right) + \Omega \cos \left( \omega_{d} t\right) \left( \hat{a}^{\dagger} + \hat{a}  \right),
\end{eqnarray}	
describing a single mode of an optical oscillator, frequency $\omega$ and creation (annihilation) operators $\hat{a}^{\dagger}$ ($\hat{a}$), interacting with a single mode of a mechanical oscillator, frequency $\omega_{m}$ and creation (annihilation) operators $\hat{b}^{\dagger}$ ($\hat{b}$), with optomechanical coupling strength $g_{0}$, and external driving of the optical mode with strength $\Omega$ and  frequency $\omega_{d}$.
For the sake of simplicity, we assume all parameters real.
This Hamiltonian model allows the description of experimentally observed optomechanical effects such as strong coupling \cite{Groeblacher2009}, mechanical squeezing \cite{Wollman2015}, entanglement \cite{Palomaki2013,Riedinger2018}, nonreciprocal behaviour \cite{Bernier2017,Barzanjeh2017}, and non-clasical mechanical states \cite{Riedinger2016,Hong2017}, among others \cite{Weis2010,Lecocq2015,Marinkovic2018}.

From a theoretical point of view, it is possible to recast the standard driven optomechanical model into an effective Hamiltonian, 
\begin{eqnarray}
	\frac{\hat{H}_{\mathrm{eff}}}{\hbar} &=& - \frac{g_{0}^{2}}{\omega_{m }} \left( \hat{a}^{\dagger} \hat{a} \right)^{2} + \frac{\Omega}{2} e^{- \frac{\alpha^{2}}{2}} \left\{  \hat{a}^{\dagger} \left[ \sum_{p=0}^{\infty}\frac{1}{p!}  (-\alpha \hat{b}^{\dagger})^{p} ~_{1}F_{1}(- \hat{b}^{\dagger} \hat{b}; p+1; \alpha^{2} )  e^{i  ( \Delta + p \omega_{m } ) t } + \right. \right. \nonumber \\
	&&  \left. \left. +\sum_{p=1}^{\infty}\frac{1}{p!} ~_{1}F_{1}(- \hat{b}^{\dagger} \hat{b}; p+1; \alpha^{2} ) (\alpha \hat{b})^{p}  e^{i  ( \Delta - p \omega_{m } ) t } \right] +  \hat{a} \left[ \sum_{p=0}^{\infty}\frac{1}{p!} ~_{1}F_{1}(- \hat{b}^{\dagger} \hat{b}; p+1; \alpha^{2} )   \right. \right. \nonumber \\
	&& \left. \left. (-\alpha \hat{b})^{p}  e^{-i  ( \Delta + p \omega_{m } ) t } + \sum_{p=1}^{\infty}\frac{1}{p!}  (\alpha \hat{b}^{\dagger})^{p}  ~_{1}F_{1}(- \hat{b}^{\dagger} \hat{b}; p+1; \alpha^{2} ) e^{-i  ( \Delta - p \omega_{m } ) t }  \right]  \right\},  
\end{eqnarray}
where we use the confluent hypergeometric function $_{1}F_{1}(a;b;z)$, define a detuning between the optical mode and driving frequencies $\Delta = \omega - \omega_{d}$, and introduce an auxiliary dimensionless parameter $\alpha = g_{0} / \omega_{m }$.
The first term of this effective Hamiltonian is a Kerr term for the optical oscillator mode, the second term vanishes in the absence of driving and we recover the well-known fact that the standard optomechanical Hamiltonian is feasible of diagonalization \cite{Restrepo2017}.
In this form, the second term allows us to realize that the driving frequency selects the type of dominant processes in the dynamics \cite{Ventura2015}. 
Driving the system with a detuning proportional to an integer of the mechanical mode frequency, $\Delta = p \omega_{m }$ with $p = 0, 1, 2, \ldots$, yields an effective Hamiltonian for parametric conversion of order $p$ between the optical and mechanical modes with Kerr nonlinearity in the optical mode, 
\begin{eqnarray}
	\left. \frac{\hat{H}_{\mathrm{eff}}}{\hbar} \right\vert_{\Delta = p \omega_{m }} &\approx& - \frac{g_{0}^{2}}{\omega_{m }} \left( \hat{a}^{\dagger} \hat{a} \right)^{2} + \frac{\Omega}{2} e^{- \frac{\alpha^{2}}{2}} \left[ \frac{1}{p!} ~_{1}F_{1}(- \hat{b}^{\dagger} \hat{b}; p+1; \alpha^{2} )   \hat{a}^{\dagger} (\alpha \hat{b} )^{p}  \right.  \nonumber \\ 
	&& \left.  + \frac{1}{p!}  ~ \hat{a}  ( \alpha \hat{b}^{\dagger} )^{p}  ~_{1}F_{1}(- \hat{b}^{\dagger} \hat{b}; p+1; \alpha^{2} ) \right], 
\end{eqnarray}
informing us that resonant driving of the optical oscillator $p=0$ produces an effective optical driven Kerr system where the drive is modified by the optomechanical coupling strength and the excitation number of the mechanical mode \cite{Gong2009,Aldana2013,Xiong2016}.
For detuning equal to the mechanical mode frequency, $\Delta = \omega_{m }$ with $p=1$, the effective Hamiltonian takes the form of an optomechanical beam-splitter with Kerr nonlinearity in the optical mode and suggests its use for coherent state transfer between the optical and mechanical modes \cite{Ventura2019}.
For the next order, $p=2$, it becomes a second order parametric down-converter that suggest its use to generate mechanical squeezed states \cite{Wollman2015}.
Choosing a driving frequency that delivers a negative detuning $\Delta = - p \omega_{m }$ with $p=1,2,3,\ldots$ yields an effective Hamiltonian,
\begin{eqnarray}
	\left. \frac{\hat{H}_{\mathrm{eff}}}{\hbar} \right\vert_{\Delta = -p \omega_{m }} &\approx& - \frac{g_{0}^{2}}{\omega_{m }} \left( \hat{a}^{\dagger} \hat{a} \right)^{2} + \frac{\Omega}{2} e^{- \frac{\alpha^{2}}{2}} \left[ \frac{1}{p!}  \hat{a}^{\dagger} (-\alpha \hat{b}^{\dagger})^{p} ~_{1}F_{1}(- \hat{b}^{\dagger} \hat{b}; p+1; \alpha^{2} )    + \right.  \nonumber \\ 
	&& \left. \frac{1}{p!}  ~_{1}F_{1}(- \hat{b}^{\dagger} \hat{b}; p+1; \alpha^{2} )  ~\hat{a}  (-\alpha \hat{b})^{p}  \right], 
\end{eqnarray}
that, for the case of detuning equal to the negative mechanical mode frequency $\Delta = - \omega_{m }$ with $p=1$ suggests its use for two-mode squeezing. 
Of course, we may think of processes involving switching from resonant to off-resonant driving to implement more complex processes like sideband cooling \cite{Riviere2011}. 

Our interest focuses on one-dimensional photonic crystal cavities etched on nanobeams whose structure may be optimized to enhance the optomechanical coupling or decrease losses \cite{Chan2012}. 
These devices localize optical and mechanical modes around a defect in a large periodic array and sustain multiple optical and mechanical modes \cite{Chan2012,Eichenfield2009,Yu2018}; the former may be addressed separately.  
In Section \ref{sec:S2}, we introduce a finite element model for such a device, the Hamiltonian describing the individual interaction of multiple optical modes with one mechanical mode, and the effective Hamiltonian form that allows us to create insight of its dynamics.
In Section \ref{sec:S3}, we propose a driving scheme that allows us to simulate a quantum parametric process where the annihilation (creation) of an optical excitation produces the first and second order creation (annihilation) of a mechanical excitation.
This parametric process in the strong driving regime, mean optical field approximation, produces an effective Hamiltonian of a parametric oscillator in the mechanical mode.
In principle, the full optical control of the parameter values in the parametric oscillator should allow probing the controlled squeezing of the mechanical mode.
We close our contribution with a brief conclusion in Section \ref{sec:S4}.

\section{Driven polychrome optomechanical model} \label{sec:S2}

\begin{figure}
	\centering
	\includegraphics{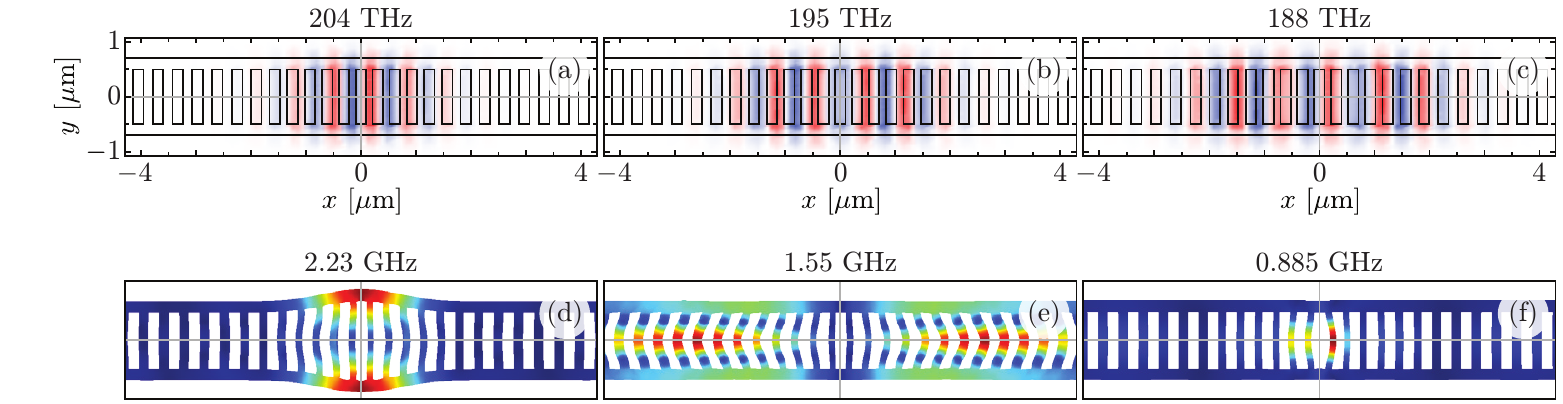}
	\caption{(a)--(c) Optical and (d)--(f) mechanical modes of a nanobeam with their corresponding frequencies. The optical modes are in the infrared telecommunication band.}\label{Fig1}
\end{figure}

Nanobeams with engraved one-dimensional photonic crystal cavities, Fig.~\ref{Fig1}, are a standard experimental realization for current optomechanical setups \cite{Chan2012, Eichenfield2009}. 
A typical device is made of a silica nanobeam where a one-dimensional photonic cavity is created by introducing a defect in an otherwise periodic structure to allow for localized optical and mechanical modes in it. 
We focus on the structure proposed in Ref.~\cite{Eichenfield2009}. 
It consists of 75 rectangular cells in a periodic array along the $x$ axis. 
Quadratic reduction of the size along the $x$-axis of 15 cells in the middle of the array introduces a defect. 
The resulting photonic defect cavity supports optical modes that approximately follow the eigenstates of the one-dimensional quantum mechanical harmonic oscillator along the $x$ axis. 
Each regular cell has length $360~\textrm{nm}$ ($x$-axis), width $1400~\textrm{nm}$ ($y$-axis), and thickness $220~\textrm{nm}$ ($z$-axis). 
These type of structures allow for many optical modes \cite{Buckley2014} which may even be coupled among themselves \cite{Yu2018}. 
We use Finite Element Modeling to find three optical modes where the electric field is mostly polarized along the $y$-axis. 
These modes have frequencies $204~\mathrm{THz}$, $195~\mathrm{THz}$ and $188~\mathrm{THz}$, Fig.~\ref{Fig1}(a), Fig.~\ref{Fig1}(b) and Fig.~\ref{Fig1}(c), in that order. 
We also find three mechanical modes localized around the defect. 
These are breathing, accordion, and pinch modes with frequency $2.23~\mathrm{GHz}$, $1.55~\mathrm{GHz}$ and $0.885~\mathrm{GHz}$, Fig.~\ref{Fig1}(d), Fig.~\ref{Fig1}(e) and Fig.~\ref{Fig1}(f), in that order.
Our results, fundamental frequencies and mode shapes, are in agreement with those of Ref.~\cite{Eichenfield2009}. 
While this structure is designed to increase the total optomechanical coupling, there is one optomechanical coupling between each pair of optical and mechanical modes and also between pairs of optical modes. All of these couplings, of course, depend on the particular details and symmetries of the structure. Finite Element techniques offer a powerful tool to study and optimize these structures \cite{Chan2012,Eichenfield2009,Buckley2014,Deotare2009}. 
Due to their difference in frequency, some of these optical modes may be separately addressed and, therefore, it becomes worthwhile to study the quantum dynamics of driven polychrome optomechanical systems.

In the following, we assume that only the optical modes are driven by an external coherent source such that the effective Hamiltonian becomes,
\begin{eqnarray}
	\frac{\hat{H}}{\hbar} = \sum_{j} \omega_{j} \hat{a}_{j}^{\dagger} \hat{a}_{j} + \omega_{m } \hat{b}^{\dagger} \hat{b} - \sum_{j} g_{0j} \hat{a}_{j}^{\dagger} \hat{a}_{j} \left( \hat{b}^{\dagger} + \hat{b}  \right) + \sum_{j} \Omega_{j} \cos \left( \omega_{dj} t\right) \left( \hat{a}_{j}^{\dagger} + \hat{a}_{j}  \right),
\end{eqnarray}	
where the optical modes, frequency $\omega_{j}$ and creation (annihilation) operators $\hat{a}_{j}^{\dagger}$ ($\hat{a}_{j}$), interact with a single mechanical mode, frequency $\omega_{m}$ and creation (annihilation) operators $\hat{b}^{\dagger}$ ($\hat{b}$), under coupling strength $g_{0j}$, external optical driving strength $\Omega_{j}$, and driving frequency $\omega_{dj}$.

We follow an approach similar to that in Ref. \cite{Ventura2015}. 
First, we move into a frame defined by the driving frequencies and the excitation number of each optical mode. 
Then, we perform an optical rotating wave approximation.
At this point, we move into a new frame defined by the displacement of the mechanical mode by a factor proportional to the optomechanical coupling and the excitation number of the corresponding optical mode divided by the mechanical frequency. 
Then, we move into a frame defined by the free optical and mechanical oscillators to obtain an effective Hamiltonian, 
\begin{eqnarray}
	\frac{\hat{H}_{\mathrm{eff}}}{\hbar} &=& -\sum_{j,k} \frac{g_{0j} g_{0k}}{\omega_{m }} \hat{a}_{j}^{\dagger} \hat{a}_{j} \hat{a}_{k}^{\dagger} \hat{a}_{k} + \sum_{j}   \frac{ \Omega_{j}}{2}  e^{- \frac{\alpha_{j}^{2}}{2}} \left\{ \hat{a}_{j}^{\dagger}  \left[  \sum_{p_{j}=0}^{\infty} \frac{1}{p_{j}!} (-\alpha_{j} \hat{b}^{\dagger})^{p_{j}} ~_{1}F_{1}(- \hat{b}^{\dagger} \hat{b}; p_{j}+1; \alpha_{j}^{2} )   \right. \right. \nonumber \\
	&&  \left. \left.   e^{i  ( \Delta_{j} + p_{j} \omega_{m } ) t } + \sum_{p_{j}=1}^{\infty}\frac{1}{p_{j}!} ~_{1}F_{1}(- \hat{b}^{\dagger} \hat{b}; p_{j}+1; \alpha_{j}^{2} ) (\alpha_{j} \hat{b})^{p_{j}}  e^{i  ( \Delta_{j} - p_{j} \omega_{m } ) t } \right]   + \mathrm{h.c.} \right\}
\end{eqnarray}	
written in terms of the optical detunings $\Delta_{j}=\omega_{j}-\omega_{dj}$ and the displacement parameters  $\alpha_{j} = g_{0j} / \omega_{m }$.
Again, we are assuming all parameters real for the sake of simplicity.
The first term of this effective Hamiltonian is a collection of self- and cross-Kerr  interactions, $\left( \hat{a}_{j}^{\dagger} \hat{a}_{j}\right)^{2}$ and $\hat{a}_{j}^{\dagger} \hat{a}_{j} \hat{a}_{k}^{\dagger} \hat{a}_{k}$, in that order. 
The second term is identical to that in the driven standard optomechanical model for each optical mode and informs us of the ability to select effective dynamics depending on the optical detuning between optical oscillators and driving fields. 
In the following, we will explore a particular driving scheme that may help us realize the uses of this polychrome optomechanical platform as an analog quantum simulator.

\section{Analog simulation of a parametric oscillator} \label{sec:S3}

We may choose to drive just two optical modes, for example, with a driving strength  that adiabatically changes with time $\Omega_{j}(t)$, close to the first and second sideband transition of the mechanical mode by a small factor that changes adiabatically in time, $p_{1} = -\omega_{m } -\epsilon(t)$ and $p_{2} = - 2 \omega_{m } - 2 \epsilon(t)$ with $\epsilon(t) \ll \omega_{m }$, such that we may approximate the effective dynamics,
\begin{eqnarray}
	\frac{\hat{H}_{\mathrm{eff}}}{\hbar} &\approx& -\sum_{j,k} \frac{g_{0j} g_{0k}}{\omega_{m }} \hat{a}_{j}^{\dagger} \hat{a}_{j} \hat{a}_{k}^{\dagger} \hat{a}_{k} + \epsilon(t) \hat{b}^{\dagger} \hat{b}  + \frac{1}{2} \sum_{j=1}^{2} \frac{\Omega_{j}(t)}{j!}  e^{- \frac{\alpha_{j}^{2}}{2}} \left[   \hat{a}_{j}^{\dagger} ~_{1}F_{1}(- \hat{b}^{\dagger} \hat{b}; j+1; \alpha_{j}^{2} ) (\alpha_{j} \hat{b})^{j}  \right. \nonumber \\
	&& \left. + \hat{a}_{j} (\alpha_{j} \hat{b}^{\dagger})^{j} ~_{1}F_{1}(- \hat{b}^{\dagger} \hat{b}; j+1; \alpha_{j}^{2} )  \right] , 
\end{eqnarray}
as long as the driving strength and the frequency change slowly in time, which is experimentally plausible in principle.

We now turn to the fact that one-dimensional photonic cavities on nanobeams have optical losses.
This produces dynamics that reaches a coherent state with large photon population as a steady state of the optical modes.
Experimental values for the losses in these devices lead to optical steady states with up to a few hundred photons; for example, the optomechanical device in Ref.~\cite{Chan2012} has losses of about $1.34~\mathrm{GHz}$ and a steady state with up to a mean optical excitation number of $|\beta|^{2}=120$ photons.
Thus, it is possible to work out an effective mean-field approximation on this driven system \cite{Ventura2019,Jaramillo2020}, such that we obtain an effective mean-field Hamiltonian,
\begin{eqnarray}
	\frac{\hat{H}_{\mathrm{mf}}}{\hbar} &\approx&  \epsilon(t) \hat{b}^{\dagger} \hat{b} + g_{1}(t)  \left( \hat{b}^{\dagger} +  \hat{b} \right) +  g_{2}(t) \left( \hat{b}^{\dagger2} + \hat{b}^{2} \right), 
\end{eqnarray}
that is identical in shape to that of a parametric oscillator, up to a constant phase, with time-dependent parameters controlled by the optical driving fields, 
\begin{eqnarray}
	g_{j}(t) &\approx& \frac{\Omega_{j}(t)}{j!} e^{- \frac{\alpha_{j}^{2}}{2}} \alpha_{j} \beta_{j},
\end{eqnarray}
where we safely assume that $ ~_{1}F_{1}(- \hat{b}^{\dagger} \hat{b}; j+1; \alpha_{j}^{2} ) \approx 1$ for typical values of the optomechanical coupling.

\begin{figure}
	\centering
	\includegraphics{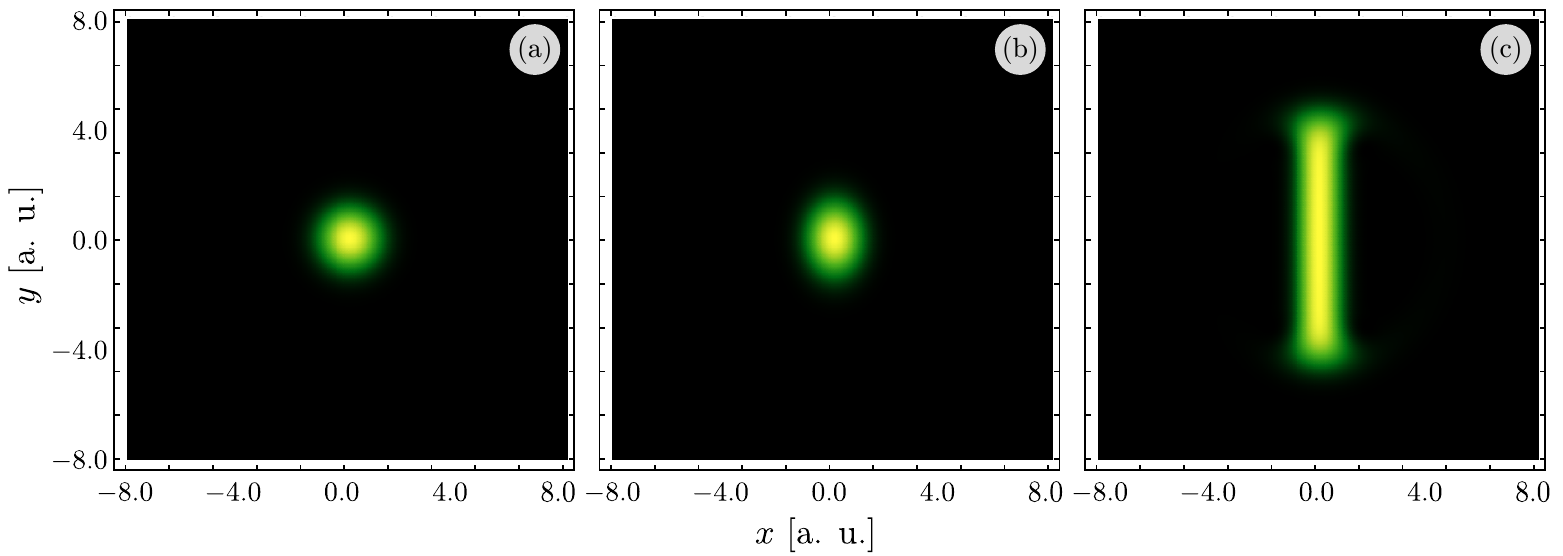}
	\caption{ Hussimi Q-function for the ground state of a parametric oscillator with parameter values $\epsilon = 10 g_{1}$ and (a) $g_{2} = 0 g_{1}$, (b) $g_{2} = 2.5 g_{1}$, and (c) $g_{2} = 0.99 \epsilon/2$, where the transition from coherent to squeezed to an eigenstate of the $\hat{x}$ canonical variable expected from the parametric oscillator may be inferred.} \label{Fig2}
\end{figure}

For the sake of providing an example, let us work out the simplest scenario where the driving strength and frequency are constant.
This allows us to move into a displaced reference frame $\hat{D}(\xi)$ with displacement $\xi=-g_{1}/(\epsilon+2g_{2})$ where the effective displaced Hamiltonian, 
\begin{eqnarray}
	\frac{\hat{H}_{\mathrm{mf}}}{\hbar} &\approx&  \frac{1}{2}\left(  \epsilon + 2g_{2} \right) \hat{x}^{2}  +\frac{1}{2} \left( \epsilon - 2g_{2} \right) \hat{y}^{2},
\end{eqnarray}
up to a constant, in terms of the scaled canonical variables ($\hat{x},\hat{y}$) defined by the relation $\hat{b}=  \left(\hat{x} + i\hat{y} \right) / \sqrt{2}$ and $\hat{b}^{\dagger}=  \left(\hat{x} - i\hat{y}\right)/\sqrt{2}$.
The effective Hamiltonian looks like that of a particle under the action of a quadratic potential $\frac{1}{2} \left( \epsilon - 2g_{2} \right)\hat{y}^{2}$ controlled by the second-order parametric driving strength. 
It takes the form of a harmonic oscillator Hamiltonian for $ \epsilon > 2g_{2}$, a free particle for $ \epsilon = 2g_{2}$, and an inverted quadratic potential for $ \epsilon < 2g_{2}$.
In principle, we may probe the model near the transition from an harmonic oscillator to a free particle, $2g_{2}  \gtrapprox \epsilon $.
For example, we may observe the deformation of the ground state from a coherent state, Fig. \ref{Fig2}(a), to a squeezed state, Fig. \ref{Fig2}(b), and to a squeezed state that starts resembling an eigenstate of the scaled canonical variable $\hat{x}$, Fig. \ref{Fig2}(c).

\section{Conclusion} \label{sec:S4}

We presented a finite element simulation of a common experimental realization of the standard optomechanical model in the form of a one-dimensional photonic cavity etched on a silica nanobeam. 
The model shows at least three optical modes in the infrared telecommunication band and three mechanical vibration modes in the GHz range localized near the photonic cavity. 
In principle, each one of the optical modes may be coherently driven on its own. 
This motivated us to explore a Hamiltonian model describing the interaction of multiple optical modes with a single mechanical mode to obtain an effective Hamiltonian producing insight on the parametric processes that may be supported by the system. 
We use this insight to propose a bichromatic driving scheme that provides us with an analog simulation of the parametric oscillator in the lossy optical cavity regime under a mean field approximation.   
The parametric oscillator is optically controlled and requires adiabatic time-dependent driving strength and frequency in its most general realization.
As a simple example, we studied the ground state of the time-independent parametric oscillator and show that it should be possible to explore the transition from a coherent state to a eigenstate of a canonical variable in the ground state of the system.

	\section*{Acknowledgments}	
	F.~E.~O., F~.H~.M.-V. and B.~R.~J.-A. thank CONACYT for their financial support.
	F.~E.~O. thanks U.~N.~N. for study leave support.  
	B.~M.~R.-L. thanks the Quantum Fest 2021 organizers for their hospitality. 

	

%
		
\end{document}